\documentclass[11pt,a4paper]{article}
\pdfoutput=1
\newcommand{\be}{\begin{equation}}
\newcommand{\ee}{\end{equation}}
\newcommand{\bea}{\begin{eqnarray}}
\newcommand{\eea}{\end{eqnarray}}
\newcommand{\beas}{\begin{eqnarray*}}
\newcommand{\eeas}{\end{eqnarray*}}
\newcommand{\ba}{\begin{array}}
\newcommand{\ea}{\end{array}}
\newcommand{\nn}{\nonumber}

\newcommand{\bt}{\begin{table}}

\newcommand{\al}{\alpha}
\newcommand{\ga}{\gamma}

\newcommand{\Ga}{\Gamma}	

\newcommand{\de}{\delta}
\newcommand{\De}{\Delta}
\newcommand{\ka}{\kappa}
\newcommand{\la}{\lambda}
\newcommand{\La}{\Lambda}
\newcommand{\na}{\nabla}

\newcommand{\si}{\sigma} 
\newcommand{\Si}{\Sigma}

\begin{document}
\title{\bf Quartet-metric gravity and dark components 
of the Universe}
\author{Yuriy~F.~Pirogov\footnote{E-mail: pirogov@ihep.ru }\\
\small{\em 
Theory Division,  
SRC  Institute for High Energy Physics
of NRC Kurchatov Institute,}\\
\small{\em Protvino,  Russia}
}
\date{}
\maketitle

\begin{abstract}
\noindent
In the report  there are presented the general  frameworks for the 
 quartet-metric gravity  based upon the  two physical concepts. 
 First, there  exist   in space-time 
the  distinct   dynamical  coordinates, given by  a scalar quartet,
playing the role of the Higgs  fields for gravity. Second,     
the  physical gravity fields arising due to the spontaneous symmetry  breaking  
serve  as the  dark components of the Universe.
It is argued that the mere admixture to metric of the scalar quartet  
may give rise to  an extremely wide spectrum of 
the emergent   gravity phenomena beyond General Relativity (GR).
Developing  the  proposed frameworks further to 
find out the next-to-GR  theory of gravity is a challenge.\\[1ex]
\end{abstract}

\section{Introduction: dark components and quartet-metric gravity}

The current  state-of-the-art in cosmology 
may be  summarized  by the so-called 
Standard Cosmological   Model (or, otherwise, 
the $\La$CDM model).\footnote{See, e.g.\ Ref.~\cite{PDG16}.} 
According to the model, the Universe expands 
with acceleration, is  spatially flat ($K=0$)  
with high precision,   and  possesses   
at  the present  epoch by the following 

\begin{itemize}

\item  {\it Energy content\/}: 
$\Omega_K=0$, $\Omega_r+\Omega_b+\Omega_{DM}
+\Omega_{\La}=1$,  with  the  partial contributions:

\begin{itemize}

\item  Radiation  negligibly small $\Omega_r\ll 1$;

\item   Baryonic matter $\Omega_b\simeq 0.04 $;

\item  Cold dark matter (CDM) $\Omega_{DM}\simeq 0.26 $ 
(the total matter $\Omega_{b+DM} \simeq 0.3$);

\item  Vacuum energy/$\La$-term  $\Omega_{\La}\simeq 0.7 $.

\end{itemize}

\end{itemize}
The $\La$CDM model  proves to be  highly successful phenomenologically 
 revealing   nevertheless  some  (mainly theoretical)

\begin{itemize}

\item {\it Problems\/}:

\begin{itemize}

\item Nature and properties of DM;

\item  Cosmological constant (CC)/$\La$-term   (un)naturalness:   
 why $\La$ is extremely   small compared to   the  Planck mass  $M_{Pl}$
 (or even  the electroweak   scale $M_W$)?

\end{itemize}

\end{itemize}
The solution to the problems  may imply, among other things, going 
beyond  General Relativity (GR), upon which $\La$CDM is  based.
In the absence of the convincing    observational and/or  
theoretical arguments in favor of a preferred  
choice for the  beyond  GR theory of gravity (if any),  it 
it is natural to try and  modify  GR  
restricting himself  exclusively by the  
gravity components  already present 
in metric.\footnote{In this respect, one may invoke   
the principle of the so-called 
Occam's razor: ``Among the competing hypotheses, 
the one(s) with the fewest assumptions should be preferred''.}
In GR, the  metric components  beyond the tensor graviton get  
non-physical  due to the general relativity, or, technically, 
the  general diffeomorphism (GDiff) invariance.
To lift these  gauge  components   from 
the gauge ``cellar'' to  the observational  ``attic'', converting them 
into  the physical dark components (DCs),
the mechanism  of the spontaneous symmetry breaking (SSB) 
for gravity is to be implemented. To this end, 
in~Ref.~\cite{Pir16} there were developed 
the frameworks of the  spontaneously broken GR,  
with the distinct dynamical coordinates, given by  a scalar quartet,  
playing the role of   the Higgs-like  fields for gravity.
The so-constructed quartet-metric  gravity  (QMG)  describes, in  general, 
the scalar-vector-tensor gravity (or some of its reductions).
At that, the physical  gravity components  arising  due to SSB are to   be  treated  
as the gravitational DM and/or DE,   depending on the context~\cite{Pir12,Pir16a}.
The respective    frameworks, with some elaboration, 
are exposed in what follows.

\section{Quartet-metric gravity: general frameworks}

\subsection{Basic ingredients}  

Whatever the  underlying theory of 
gravity might be, to be eventually confronted with Particle Standard Model, 
as well as the observations,  it should be realized 
at  the present-day level (or even  at  the level of a foreseeable future) 
as  the effective field theory (EFT).
The latter   is generically characterized  by the three basic ingredients: 
the set  of fields, the pattern of  symmetries 
and the type of  interactions among the given fields
satisfying the given symmetries.  QMG  
 may be  characterized  as follows.  

\begin{itemize}

\item {\it Field set\/}:

\begin{itemize}

\item Gravity field/metric $g_{\mu\nu}(x)$;

\item  Scalar quartet   
  ${{Z}}^a(x), \ \ a=0,\dots3,$ 
  defining  on  space-time the distinct  (chart-vise  reversible) dynamical  coordinates 
 ${z}^\al =\de^\al_a {{Z}}^a(x)$, $x^\mu= x^\mu ({z}) $,  
 with $x^\mu$ the arbitrary observer's coordinates.   By default, 
 such distinct {\it quasi-affine}  coordinates $z^\al$  are  attributed to  
the (chart-wise homogeneous) vacuum.  

 \end{itemize}

 \item {\it Symmetry pattern\/}:

 \begin{itemize}
 
    \item  General covariance (GC):
    all the (smooth enough) coordinates 
    $x^\mu$ (including, in particular, ${z}^\al$) 
    on the space-time manifold are {\it a priori} admissible 
    {\em on par\/}.\footnote{Paraphrasing the well-known maxim
    one may thus say that  ``all  coordinates are equivalent 
    but some are more equivalent than others''.}

    \item General relativity (GR)/general diffeomorphism  (GDiff) invariance under 
  \be
  {\rm GDiff} \ : \ x^\mu \to x^\mu - \xi^\mu(x), \  \  |\xi^\mu|\ll 1,
   \ee
   followed by the associated   transformation 
   of the matter fields through their Lie derivative. 
   At that, GDiff  serves as the gauge group for gravity.
   
   \item  Poincare invariance:
  \be
  ISO(1,3) \ : \ {{Z}}^a \to \La^a{}_b {{Z}}^b +c^a,
  \ee
 with  the   (chart-wise) global  translations $c^a$ 
 and  Lorentz transformations   $\La$ living invariant   
 the   Minkowski symbol  $\eta_{ab}$.

 \end{itemize}
 
 \item     {\it Type of interactions\/}:  defined  generically by  
the gauge and Lorentz-invariant  GC scalar action
\be
      S= \int {\cal L}(g_{\mu\nu}, \partial_\la       g_{\mu\nu},   {{Z}}^a_\mu,  
       \partial_\mu {{Z}}^a_\nu,       \Psi,\partial_\mu \Psi, \eta_{ab}) d^4x,
      \ee
with $\cal L$ the  Lagrangian density   in terms of 
the   tetrad ${Z}^a_{\mu}\equiv \partial_\mu {{Z}}^a$ and  
a generic matter field $\Psi$.\footnote{Due to the assumed
(chart-wise) global Poincare invariance,  reflecting the homogeneity of the vacuum, 
there is no derivativeless input of  ${{Z}}^a$. This allows   
to treat the second derivative of $Z^a$  
just as the first derivative of the tetrad ${{Z}}^a_{\mu}$.}  

\end{itemize}

\subsection{Basic constructions} 

The looked-for GC scalar    action may  straightforwardly  be constructed in terms of the 
GC elements as follows.

 \begin{itemize}
    
    \item {\it Connection-like  tensor\/}:~\cite{Pir16}
 \be
 B^\la_{\mu\nu}(x)\equiv   \Ga^\la_{\mu\nu}(g_{\mu\nu}) 
-\partial_\mu\partial_\nu z^\al \partial x^\la/\partial z^\al|_{z=Z(x)},
 \ee
where $z^\al\equiv \de^\al_a Z^a(x)$
and,  inversely, $x^\la= x^\la(z)$. 
At that, in the  quasi-affine coordinates $z^\al$ one gets 
$B^\ga_{\al\beta}(z)= \Ga^\ga_{\al\beta}(g_{\al\beta})$.

 \item {\it Quasi-affine metric\/}:
\be  
   \zeta_{\mu\nu}={Z}^a_{\mu}{Z}^b_{\nu}\eta_{ab},
   \ee
in terms of which one gets,  in particular,
 \be
B^\la_{\mu\nu}=  \Ga^\la_{\mu\nu}( g_{\mu\nu}) 
- \ga^\la_{\mu\nu}(\zeta_{\mu\nu}) 
 \ee
 as the difference of the two Christoffel  
 connections.\footnote{It follows  that in 
 the quasi-affine coordinates $z^\al$ 
one has $\zeta_{\al\beta}(z)=\eta_{\al\beta}$, $\zeta=-1$   
and  $\ga^\ga_{\al\beta}(z)=0$, what, in fact,  distinguishes  such coordinates.} 

\item  {\it Scalar-graviton\/}: 
\be
    \si\equiv  1/2\,\ln g/\zeta	
    \ee
composed of the two scalar densities 
    $g\equiv \det (g_{\mu\nu})$ \  and \ $\zeta\equiv \det (\zeta_{\mu\nu})$. 
 
 \item{\it Higgs-like scalar\/}:\footnote{Such a scalar to implement 
 the Higgs  mechanism for  graviton 
was proposed originally in Refs.~\cite{Muh, Oda1,Oda2}.}
\be
\Si^{ab}\equiv g^{\ka\la} {Z}^a_{\ka} {Z}^b_{\la}, 
\ee
or, equivalently,  the {\it Higgs-like tensor}
\be
 \Si^\mu{}_\nu  \equiv   {Z}^{\mu}_a\Si^{ac}\eta_{cb}
{Z}^b_{\nu}=  g^{\mu\la} \zeta_{\la\nu} ,
 \ee
with ${Z}^{\mu}_a\equiv \partial x^\mu/\partial  z^\al|_{z=Z(x)}$    the tetrad  inverse to ${Z}^a_{\mu}$.  
This implies  the relation
\be
  \det( \Si^{ac}\eta_{cb})=    \det( \Si^\mu{}_\nu)=\zeta/g =e^{-2\si}
\ee
and defines another   scalar  composed  of the two basic metrics:
\be
s_1 \equiv  \Si^{ab}\eta_{ba} =\Si^\mu{}_\mu=   g^{\mu\nu} \zeta_{\nu\mu} ,
\ee
as well as   similar constructions of the higher 
degrees, like $s_2=\Si^\mu{}_\nu \Si^\nu{}_\mu\neq s_1^2$, etc.

 \item {\it  Space-time measure\/} $\cal M$ for the 
volume element $d V\equiv{\cal M}d^4 x$:

\begin{itemize}

\item Gravitational:
${\cal  M}_g=\sqrt{-\det (g_{\mu\nu})} \equiv \sqrt{-g} $; 

\item  Quasi-affine/non-gravitating:\footnote{The  quartet 
of  scalar fields for the dynamical non-gravitating 
measure was proposed originally  in Ref.~\cite{Guendel96}. 
For a  subsequent elaboration, see, e.g.~\cite{Guendel08}.} 
$  {\cal  M}_\zeta=  \det (\partial_\mu 
{{Z}}^a)= \sqrt{-\det (\zeta_{\mu\nu})}\equiv  \sqrt{-\zeta}$.

\end{itemize}
\end{itemize}
All the constructions    are understood ultimately as  the functions of 
the basic fields $g_{\mu\nu}$ and $Z^a_\mu=\partial_\mu Z^a$.
In building  the action,  one should, in particular,  account for  

\begin{itemize}

\item {\it Measure/Lagrangian ambiguity\/}: 
without loss of generality one can always 
choose either the gravitational or quasi-affine  measure  (or a combination of both)
supplemented by a proper redefinition of the Lagrangian, which one gets  simpler. 

\end{itemize}
Imposing on space-time  the geometrical structure
one gets   more elaborate  elements. 

\begin{itemize}

\item  {\it Quartet-modified metric\/}: 
\be
\tilde g_{\mu\nu}  = f(\Si)  g_{\mu\nu} +  \varphi(\Si)  \zeta_{\mu\nu}
\ee
as  a   linear combination  of the two basic metrics, 
with the $\Si$-dependent coefficients.
This defines, in turn,  the quartet-modified   Higgs-like tensor
\be
\tilde \Si^\mu{}\nu  \equiv \tilde g^{\mu\la}\zeta_{\la\nu}=
f(\Si)  \Si^\mu{}_\nu +  \varphi(\Si)  \de^\mu_\nu,
\ee
the  connection-like tensor
 \be
\tilde  B^\la_{\mu\nu}\equiv   \Ga^\la_{\mu\nu}(\tilde g_{\mu\nu}) 
- \ga^\la_{\mu\nu}(\zeta_{\mu\nu})
 \ee
and the   measure  
\be
\tilde {\cal  M}= \sqrt{-\det (\tilde g_{\mu\nu})}\equiv \sqrt{-\tilde g} .
\ee

\end{itemize}
{\it A priori\/}, it is  conceivable using  in the action $S$ the  different  metrics.
To reduce such an  ambiguity one may  additionally impose

\begin{itemize}

\item 
{\it Geometry universality\/}: all the geometrical constructions 
on the space-time, such as 
the Christoffel connection $\tilde \Ga^\la_{\mu\nu}$, 
 the covariant derivative 
$\tilde \na_\mu$, the measure  $\tilde {\cal M}=\sqrt{-\tilde g}$, etc,  
are to be  attributed to the single metric $\tilde g_ {\mu\nu}$. 
At  $\tilde g_ {\mu\nu} \neq  g_ {\mu\nu}$ the 
geometry/gravity equivalence is  violated due to  influence  
of  the vacuum  on metric.

\end{itemize}
Under the chosen   {\it geometry}, the residual freedom  in the action $S$ is to be attributed 
to  the   choice of  the  Lagrangian~$L$ specifying  the  {\it dynamics}.
By this token one gets

\begin{itemize}

\item  {\it GC  scalar action\/}
\be
     S= \int L(\tilde R,\tilde B^\la_{\mu\nu}, \tilde g_{\mu\nu},   
     \tilde\Si^\mu{}_\nu, \tilde  \na_\mu \Psi, \Psi ) \tilde {\cal  M} d^4x,
     \ee
   with   
   $ \tilde R\equiv R(\tilde g_{\mu\nu})$  the Ricci scalar,
   and  the Lagrangian 
   $L=K-V +L_m$ containing

\begin{itemize}

\item Kinetic term:   
$K(\tilde R, \tilde B^\la_{\mu\nu}, \tilde g_{\mu\nu},\tilde\Si)$;

\item Potential:   $V (\tilde\Si) $;

\item  Matter Lagrangian:   
$L_m(  \tilde\na_\mu \Psi, \Psi, \tilde g_{\mu\nu}, \tilde\Si)$.

\end{itemize}

\end{itemize}
Admitting in $L$ conventionally  not higher then the first degree of 
derivative of $g_{\mu\nu}$ and $Z^a_\mu$
one may,  following  the simplicity spirit,  impose  on $S$ additionally 
the reduction of the two following principle  types. 

\begin{itemize}

\item
{\it Quadratic-derivative\/}: leaving not higher than the second 
power of the first derivative of $g_{\mu\nu}$ and 
$Z^a_\mu$, but  {\em a~priori\/} 
 any  input  of $\zeta_{\mu\nu}$ (including those through~$\zeta$);

 \item
{\it Scalar-graviton\/}: leaving  the input   
of $\zeta_{\mu\nu}$ only through its  determinant $\zeta$, 
but  {\em a~priori\/}  any power of the  first derivative of $\si$. 

\end{itemize}
Imposing  both types of  reduction for the scalar graviton
results in the maximal simplification  (see, below).

\section{Quartet-modified GR: quadratic-derivative reduction}

\subsection{Non-linear theory}  

To illustrate the general frameworks let us consider in more detail 
the quartet-modified GR~\cite{Pir16}, which still possesses by 
the  geometry/gravity equivalence as   GR, 
 $\tilde g_{\mu\nu}= g_{\mu\nu}$ and $ {\cal M}=\sqrt{-g}$,
though  with   the  modified  quadratic-derivative  action: 
\be
S=\int\Big(L_g(g_{\mu\nu})  + \De L_g(g_{\mu\nu}, \zeta_{\mu\nu} ) 
+L_m(\Psi, g_{\mu\nu},  \zeta_{\mu\nu})\Big)\sqrt{-g}\,d^4 x.
\ee
In the above,
\be
L_g=-\frac{1}{2}\ka_g^{2}R(g_{\mu\nu})
\ee
is the GR Lagrangian, $\ka_g$ the Planck mass, and
\be
\De  L_g =\De K (B^\la_{\mu\nu}, g_{\mu\nu},\Si^\mu{}_\nu) -V(\Si^\mu{}_\nu)
\ee
is the additional gravity  Lagrangian  consisting of  
the potential $V$, containing the mass  and  self-interaction terms, 
and the quadratic kinetic term $\De K$.  
The latter  is given by the sum,  $\De K=\sum \ka_i  K_i$,  of 
 the partial   contributions as follows:
\bea
 {K}_1= g ^{\mu\nu} B_{\mu\ka}^\ka B_{\nu\la}^\la,   && 
 {K}_2=g_{\mu\nu}  g^{\ka\la}g^{\rho\si}
B_{\ka\la}^\mu   B_{\rho\si}^\nu,\nn\\
 {K}_{3}=g^{\mu\nu} B_{\mu\nu}^\ka B_{\ka\la}^\la,      &&  
{K}_4 = g_{\mu\nu}  g^{\ka\la}  g^{\rho\si} B_{\ka\rho}^\mu
B_{\la\si}^\nu    ,\nn\\  
 {K}_5=  g^{\mu\nu}  B_{\mu\ka}^\la  B_{\nu\la}^\ka,   && 
\eea
with the coefficients   $\ka_i$, generally,  
dependent on  $\Si$.

\subsection{Linear approximation: physics content}

Choosing  the flat backgrounds, putting 
$x^\mu =\de^\mu_\al {z}^\al$, designating ${Z}^\al=\de_a^\al {Z}^a$,  
and  (with indices operated by $\eta_{\al\beta}$)  decomposing the fields as
\bea
{Z}^\al&=& {z}^\al+ \zeta^\al,  \nn\\
g_{\al\beta}&=& \eta_{\al\beta}+h_{\al\beta},\nn\\
\zeta_{\al\beta}&=&\eta_{\al\beta}+(\partial_\al
\zeta_\beta+\partial_\beta\zeta_\al),  \nn\\
\Si_{\al\beta}&=&\eta_{\al\beta}-h_{\al\beta}+(\partial_\al
\zeta_\beta+\partial_\beta\zeta_\al),  
\eea
one arrives   in the linear approximation (LA)  at the  substitutions
\bea
\mbox{\rm LA} \ : \  \  h_{\al\beta} \to  \chi_{\al\beta}&=&h_{\al\beta} -(\partial_\al
\zeta_\beta+\partial_\beta\zeta_\al),\nn\\
h\equiv h_\al^\al \to\chi&=&h -2\partial_\al \zeta^\al.
\eea
This is nothing but  the realization  in LA of 
the SSB mechanizm for gravity.
In the  general case, GDiff gets completely broken/''hidden``, 
with   all ten gravity components 
$\chi_{\al\beta}$ becoming physical  and  describing  
the scalar, vector and tensor gravitons. 
Nevertheless,  depending on the particular choice  
for the Lagrangian parameters, there 
may be left some residual Diff symmetry to
eliminate a part of the components (especially  those 
corresponding to the putatively ''dangerous`` vector graviton)~\cite{Pir16}.
To illustrate, let us note the two following physically typical  cases.

\begin{itemize}

\item {\it  Massive scalar graviton\/}~\cite{Pir16}  $\si=\chi/2$ at 
\be
\De  K =\ka_1 K_1=\ka_1 g^{\mu\nu}\partial_\mu\si  \partial_\nu\si, 
\ee
with $\ka_1$, generally, dependent on $\si$, and  $V=V(\si)\simeq m^2_s/2\,  \si^{2}$ in addition 
to  $L_g(\chi_{\al\beta})$ of GR.
Due to   the residual (three-parameter)
transverse diffeomorphism (TDiff) symmetry
\be
{\rm TDiff} \  ; \ \chi_{\al\beta}\to   \chi_{\al\beta}+
(\partial_\al\varphi_\beta+\partial_\beta\varphi_\al), \ \ 
\partial_\ga \varphi^\ga=0,
 \ee
transforming  the trace and traceless part of $\chi_{\al\beta}$ 
separately as the proper irreducible representations, 
there is left the three physical components: 
one for the massive scalar graviton
plus two for the massless tensor one, as  
in GR.\footnote{For the scalar graviton in 
the DM and DE contexts, see~\cite{Pir12, Pir16a}. At that,
the results obtained in~\cite{Pir12} at a non-dynamical 
scalar density remain still  valid at the dynamical $\zeta$.} 

\end{itemize}

\begin{itemize}

\item {\it Massive tensor graviton\/}~\cite{Muh, Oda1,Oda2}  
$ \ga_{\al\beta}= \chi_{\al\beta}/2$    
at  $\De K=0$ and $V=V(\Si)$  chosen properly to reduce 
in LA to the  conventional Fiertz-Pauli potential  
$V_{\rm FP}(\chi_{\al\beta})$.
Under such a choice, there is left for the massive tensor graviton 
the five physical components, without  ghosts.

\end{itemize}
A wide  variety of the more sophisticated  cases  for SSB of GDiff  
 may  {\it a priori} be envisaged in QMG, with DCs beyond 
 the   two simplest realizations discussed above.
All such  cases can hardly be rejected just {\it ab initio\/}. 
To explore them  is a  challenge.

 \section{Resume}

\begin{itemize}        

\item  QMG  presents a  bold and   new type of the GR  modification 
built on  the   mechanism of SSB for gravity. 
This  is assured by  a set of the (chart-wise reversible) quasi-affine  dynamical
coordinates, given by  the scalar quartet, serving   
as the  gravity counterpart of  the  Higgs fields. 

\item 
The  mere  admixture to  metric of the scalar quartet is able to   give  rise
to the extremely  wide spectrum  of the emergent gravity 
phenomena  beyond GR, such as  the massive tensor and 
scalar  gravitons,   the gravitational DM and/or DE, etc.

\item  
Further theoretical and phenomenological study of QMG
is needed to validate  it (if any) as a candidate 
on the  beyond GR  theory of gravity.
Accounting  for the influence  of  vacuum,  QMG  as EFT  may 
ultimately  pave the way towards a more fundamental 
theory of  gravity and space-time.

\end{itemize}

\end{document}